\documentclass[manuscript,twocolumns]{aa}

\usepackage{natbib}
\bibpunct{(}{)}{;}{a}{}{,}

\newcommand{\bex}{{ XTE\,J1946$+$274}}

\usepackage{multirow}
\usepackage{graphicx}
\usepackage{amsmath}
\usepackage{amssymb}
\usepackage{txfonts}
\usepackage{epstopdf}
\usepackage{enumerate}
\usepackage{color}
\usepackage{gensymb}

\title{Detection of a large Be circumstellar disk during X-ray quiescence of XTE\,J1946$+$274}
\author{M. \"{O}zbey Arabac{\i}
     \inst{1}
    \and
    A. Camero-Arranz 
    \inst{2}
    \and
    C. Zurita
    \inst{3,4}
    \and
     J. Guti\'{e}rrez-Soto     
    \inst{5,6}
    \and
    E. Nespoli
    \inst{7}
    \and 
      J. Suso
      \inst{7}
      \and 
      F. Kiaeerad
     \inst{8,9}
      \and 
       J. Garc\'{i}a-Rojas
      \inst{3,4}
      \and
       {\"U}. K{\i}z{\i}lo{\v g}lu
      \inst{1}
      }
     \institute{$^1$ Department of Physics, Middle East Technical University, Ankara, 06531, Turkey.\\
      $^2$ Institut de Ci\`{e}ncies de l'Espai, (IEEC-CSIC), Campus UAB, Fac. de Ci\`{e}ncies, Torre C5 pa., 08193, Barcelona, Spain.\\   
      $^3$ Instituto de Astrof\'{\i}sica de Canarias, E-38200, La Laguna, Tenerife, Spain.\\     
      $^4$ Universidad de La Laguna, Dept. Astrof\'{i}sica, E-38206, La laguna, Tenerife, Spain.\\
      $^5$ Universitad de Valencia, Dept. Did\'{a}ctica de las Matem\'{a}tica,  Avda. Tarongers, 4, 46022, Valencia, Spain. \\
      $^6$ Instituto de Astrof\'{i}sica de Andaluc\'{i}a (CSIC), Glorieta de la Astronom\'{i}a s/n, 18008, Granada, Spain.\\
      $^7$ Observatorio Astron\'{o}mico de la Univ. de Valencia, C/Catedr\'{a}tico Jose Beltran, 2, 46980 Paterna (Valencia), Spain. \\
      $^8$ Nordic Optical Telescope, Apartado 474, 38700 Santa Cruz de La Palma, Spain.\\
      $^9$ Department of Astronomy, Oscar Klein Center, Stockholm University, AlbaNova, Stockholm SE-10691, Sweden.\\
}  

\authorrunning{\"{O}zbey Arabac{\i} et al.}
\titlerunning{X-ray quiescent states of XTE\,J1946$+$274}
\offprints{M. \"{O}zbey Arabac{\i}\\  \email{mehtap@astroa.physics.metu.edu.tr}}
\date{Received ; accepted}

\begin{document}

\abstract{}
	{We present a multiwavelength study of the Be/X-ray binary system XTE J1946$+$274 with the main goal 
of better characterizing its behavior during X-ray quiescence. We also aim to shed light on the possible
mechanisms which trigger the X-ray activity for this source.}
	{XTE J1946$+$274 was observed by \textit{Chandra}-ACIS during quiescence in 2013 March 12. In addition, this 
source has been monitored from the ground-based astronomical observatories of El Teide (Tenerife, Spain), Roque de los Muchachos 
(La Palma, Spain) and Sierra Nevada (Granada, Spain) since 2011 September, and from the T\"{U}B\.{I}TAK National 
Observatory (Antalya, Turkey) since 2005 April. We have performed spectral and photometric temporal analyses in order to investigate 
the quiescent state and transient behavior of this binary system.}
    {Our optical study revealed that a long mass ejection event from the Be star took place in 2006, 
lasting for about seven years, and another one is currently ongoing. We also found that a 
large Be circumstellar disk is present during quiescence, although major X-ray activity is not observed.
 We made an attempt to explain this by assuming the permanently presence of a tilted (sometimes warped) Be decretion disk. 
The 0.3--10 keV X-ray spectrum of the neutron star during quiescence was well fitted with either an absorbed black-body or 
an absorbed power-law models. The main parameters obtained for these models were kT=1.43$\pm$0.17 
and $\Gamma$=0.9$\pm$0.4 (with $N_{\rm H}\sim$2--7$\times10^{22}$ cm$^{-2}$). The 0.3--10\,keV flux of the source was 
$\sim$0.8--1$\times10^{-12}$\, erg$^{-1}$\,cm$^{-2}$s$^{-1}$. Pulsations were found 
with P$_{pulse}$=15.757(1)\,s (epoch MJD 56363.115) and an rms pulse fraction of 32.1(3)$\%$. 
The observed X-ray luminosity during quiescent periods was close to that of expected in supersonic propeller regimen.} 
{}

{}

\maketitle

\section{Introduction}

The hard X-ray transient XTE J1946$+$274 is one of the poorly-understood sources among Be/X-ray binaries 
(BeXRB), although its X-ray behavior have been 
studied in detail since its discovery with All Sky Monitor (ASM) on board the $Rossi$~$X$-$Ray$ $Timing$ 
$Explorer$ $(RXTE)$ in 1998 \citep{smtake98}. The system showed two main transient X-ray active phases 
detected with different X-ray satellites between 1998 and 2011. The first and the longest X-ray 
activity lasted about $\sim$3 years (between September 1998--August 2001) including 13 consecutive 
outbursts \citep{wilson03}. During the initial outburst of these series, having the peak X-ray flux of 
$\sim$110 mCrab in 2--60 keV band, it was revealed that the system had an X-ray pulsar with a spin period of 
15.83$\pm$0.02\,s \citep{smtake98,wilson981} orbiting around its Be companion with a period of 169.2 days in 
an 0.33 eccentric orbit \citep{campana99,wilson03}. Using optical and IR observations 
\cite{verrecchia02} found a distance for XTE J1946$+$274 in the range of 8--10 kpc, in agreement with 
the value of 9.5$\pm$2.9 kpc obtained from the X-ray data by \cite{wilson03}. In addition, 
the existence of a cyclotron resonance scattering feature (CRSF) at $\sim$35 keV
 was reported by \citet{heindl01} using the 1998 outburst observations of High Energy X-Ray Timing Experiment 
(HEXTE) and Proportional Counter Array (PCA) on \textit{RXTE}.

After a $\sim$9 years quiescence in X-rays, the system underwent a new outburst phase starting on 
2010 June 4, reaching a value of 140 mCrab in the 15--50 keV energy band within $\sim$22 days on the 
\textit{Swift}/Burst Alert Telescope (BAT) hard X-ray transient monitor \citep{krim10,muller12}. Similar to 
the previous active phase of the source, the second outburst period was again in a series including an 
initial giant outburst followed by four fainter outbursts \citep{camero101,caballero10,nakajima10,muller10}. 
The presence of another CRSF at $\sim$25 keV was discovered indicating 
the variation of cyclotron lines between the different outbursts \citep{muller12}.  

In general, X-ray active phases of XTE J1946$+$274 include 
two outbursts per orbital period that are possible to be produced if the Be disk and the 
orbital plane are offset \citep{priedholt87}. However for XTE J1946$+$274 
the outbursts do not coincide with the time of periastron and apastron passages 
of the neutron star (NS) that would be expected to happen for similar misaligned Be/X-ray systems 
\citep[and references therein]{wilson03,muller12}. Therefore some other additional mechanisms would be responsible for 
this unique behavior.

The optical/IR counterpart to XTE J1946$+$274 was discovered by \citet{verrecchia02} nearly $\sim$3 years after 
the first X-ray activity of the system. It is a relatively faint $V$ = 16.9, reddened B0--1 IV--Ve type 
star having strong H$\alpha$ and H$\beta$ emission lines in its spectra suggesting the presence of 
the decretion disk. Subsequent optical spectroscopic observations revealed the profile variations of 
H$\alpha$ emission lines implying the existence of global density perturbations in the disk \citep{wilson03}. 

In this work we present the results of a long-term multiwavelength campaign of XTE J1946$+$274. 
Since this X-ray transient spends most of its life in X-ray quiescent phase, our spectroscopic and photometric 
data mainly cover these inactive periods between the X-ray brightening phases. These optical/IR data 
come from our monitoring program on Be/X-ray binaries, which involves several ground-based astronomical 
observatories. In addition, we used one \textit{Chandra}-ACIS 
pointed observation carried out during X-ray quiescence in 2013, and several survey data from different space-borne 
telescopes to investigate the relation between the X-ray and optical/IR bands.

\begin{figure}
\hspace{-0.9cm}\includegraphics[scale=0.69, trim=3mm 4mm 1mm 0mm]{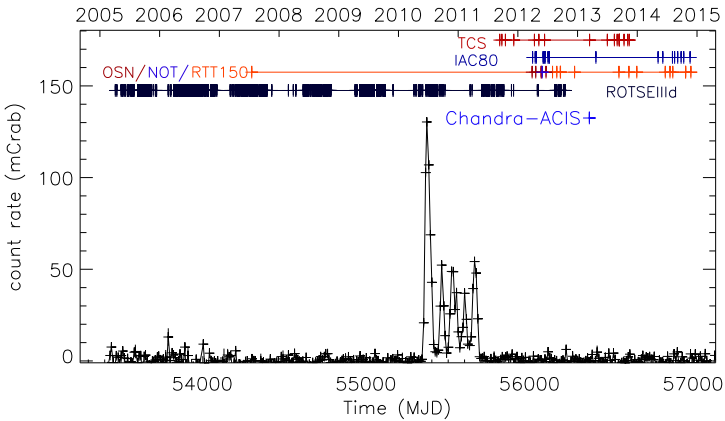}
\caption{\textit{Swift}/BAT light-curve (15--50 keV) with a bin size equal to 10\,d. Tick 
marks on the segments located above the light-curve  denote the times of the \textit{Chandra} 
pointing observation (light blue), as well as the optical/IR photometric data 
from the ground-based telescopes ROTSEIIId (black), IAC80 (dark blue) and TCS (dark red). The optical 
spectroscopic observations come from OSN (dark red), NOT (blue) and RTT150 (light red) 
(see also Table~\ref{xte_ew}).}
\label{plot_log}
\end{figure}

\begin{table}
\caption{Log of the \textit{Chandra}-ACIS observation during quiescence period II.}                          
\begin{tabular}{lllllll} 
\hline\hline 
\hspace{-0.15cm}ObsID &\hspace{-0.15cm}Date Obs. &\hspace{-0.07cm}MJD     & \hspace{-0.15cm}Exp. &\hspace{-0.15cm}Rate     	& \hspace{-0.20cm}P$_{pulse}$  &\hspace{-0.27cm} PF$^a_{rms}$	\\
                               &\hspace{-0.14cm}  (UT)         &                  &\hspace{-0.2cm} (ks)  &\hspace{-0.24cm}   (c/s)   &\hspace{-0.12cm} (s)             &\hspace{-0.21cm} ($\%$ ) \\
\hline \hline 
\hspace{-0.15cm}14646 &\hspace{-0.21cm}2013-03-12    &\hspace{-0.32cm} 56363.115 &\hspace{-0.15cm}4.6  &\hspace{-0.31cm}0.042(3) &\hspace{-0.42cm} 15.757(1)  &\hspace{-0.42cm}   32.1(2)    \\
           &\hspace{-0.32cm} 02:45:36.00    &                   &          &                     &          &      \\
\hline                        
\end{tabular}

$^a$(further details in Sect.~\ref{timing}).\\
\label{log_chandra}  
\end{table}

\begin{table}
\caption{H$\alpha$ equivalent width (EW) measurements of 
optical counterpart to XTE J1946$+$274.}
\begin{center}
\begin{tabular}{lllrc}
 \hline\hline
\hspace{-0.01cm} DATE& MJD& EW (\AA)&FWHM (\AA)&Telescope\\
\hline\hline
2007-Jul-18&54299.816&$-$37.35$\pm$1.39&9.88$\pm$0.27&RTT150\\
2012-Mar-27 &56013.148 &$-$17.73$\pm$0.76 &11.09$\pm$0.46 &OSN\\
2012-Apr-18 &56035.132 &$-$28.49$\pm$1.20 &11.57$\pm$0.30 &OSN\\
2012-May-22 &56069.014 &$-$41.65$\pm$1.31 &10.19$\pm$0.46 &OSN\\
2012-May-22 &56069.039 &$-$40.70$\pm$1.39 &9.87$\pm$0.15 &OSN\\
2012-May-28 &56075.024 &$-$45.16$\pm$1.36 &10.40$\pm$0.78 &NOT\\
2012-Jun-19 &56097.997 &$-$39.87$\pm$1.52 &10.30$\pm$0.11 &OSN\\
2012-Jul-4 &56112.963 &$-$45.07$\pm$1.01 &9.69$\pm$0.74 &OSN\\
2012-Jul-5 &56113.030 &$-$47.62$\pm$0.95 &10.21$\pm$0.94 &OSN\\ 
2012-Jul-29 &56137.922 &$-$39.12$\pm$0.86 &10.12$\pm$0.32 &RTT150\\
2012-Aug-11 &56150.981 &$-$44.46$\pm$1.18 &10.62$\pm$0.44 &OSN\\
2012-Aug-24 &56163.847 &$-$41.64$\pm$1.74 &10.14$\pm$0.31 &RTT150\\
2012-Sep-15 &56185.758 &$-$40.00$\pm$0.86 &9.78$\pm$0.31 &RTT150\\
2012-Sep-16 &56186.734 &$-$39.65$\pm$1.15 &9.76$\pm$0.31 &RTT150\\
2012-Oct-09 &56209.917 &$-$42.18$\pm$1.11 &11.66$\pm$0.26 &OSN\\
2012-Dec-11 &56272.669 &$-$42.58$\pm$2.04 &10.35$\pm$0.31 &RTT150\\
2013-Sep-08 &56543.815 &$-$40.11$\pm$1.35 &9.88$\pm$0.31 &RTT150\\
2013-Nov-08 &56604.714 &$-$36.11$\pm$0.58 &13.33$\pm$0.13 &RTT150\\
2013-Dec-26 &56652.890 &$-$38.68$\pm$1.79 &14.09$\pm$0.88 &RTT150\\
2014-Jun-19 &56827.041 &$-$39.60$\pm$0.97 &9.07$\pm$0.49 &RTT150\\
2014-Jul-17 &56855.813 &$-$42.79$\pm$1.10 &9.95$\pm$0.60 &RTT150\\
2014-Aug-03 &56872.859 &$-$41.06$\pm$0.80 &9.85$\pm$0.43 &RTT150\\
2014-Oct-20 &56950.753 & -43.14$\pm$1.28 & 9.83$\pm$0.49 &   RTT150 \\
2014-Oct-21 &56951.774 & -42.37$\pm$1.15 & 9.99$\pm$0.45 &  RTT150 \\
2014-Nov-22 &56983.714 & -38.24$\pm$1.12 & 10.00$\pm$0.47 &  RTT150 \\
\hline
\end{tabular}
\end{center}
\label{xte_ew}
\end{table}

\begin{table*}
\caption{Optical and IR magnitudes of XTE J1946$+$274 observed with IAC80 and TCS telescopes respectively.}
\begin{center}
\begin{tabular}{lcccccc}
 \hline\hline
\hspace{-0.01cm}DATE&MJD&     B            & V                & J 		 & H 		  & K$_{s}$\\
\hline\hline
2011-Sep-11   &  55815.029 & 		      &	 		 & 12.450$\pm$0.028   & ---		 &  ---      \\
2011-Sep-23   &  55827.959 & ---		& ---		   & 12.694$\pm$0.059	& 11.727$\pm$0.048 &  ---	 \\
2011-Sep-24   &  55828.962 & ---		& ---		   & -- 		& ---		   & 11.402$\pm$0.066  \\
2011-Oct-15   &  55849.970 & ---		& ---		   & 12.562$\pm$0.000	& 11.789$\pm$0.019 & 10.520$\pm$0.088  \\
2011-Dec-05   &  55900.815 & ---		& ---		   & 12.602$\pm$0.042	& 11.899$\pm$0.152 & 11.260$\pm$0.055  \\
2012-Mar-30   &  56016.195 & 18.737$\pm$0.056 & 15.784$\pm$0.050 &   ---	      & ---		 & ---\\		       
2012-Apr-10   &  56027.140 & 		      &       		 &  12.560$\pm$0.073  & 11.828$\pm$0.009 & 11.240$\pm$0.105  \\
2012-Apr-15   &  56032.192 & 18.803$\pm$0.062 & 15.89 $\pm$0.051  & --- 	      & ---\\
2012-Apr-22   &  56039.195 & ---  	      & 15.898$\pm$0.051  & --- 	      & ---\\
2012-May-07   &  56054.179 & ---              & ---	 	  & 12.734$\pm$0.083  & 11.856$\pm$0.019 & 11.326$\pm$0.136  \\
2012-Jun-02   &  56080.183 & 18.788$\pm$0.055 & 15.835$\pm$0.050  & --- 	      & ---		  & ---\\
2012-Jun-04   &  56082.206 & 18.627$\pm$0.074 & 15.784$\pm$0.051  & --- 	      & ---		  & ---\\
2012-Jun-11   &  56089.214 & 18.822$\pm$0.069 & 15.793$\pm$0.050  & --- 	      & ---		  & ---\\
2012-Jun-14   &  56092.957 & 18.708$\pm$0.063 & 15.841$\pm$0.051  & --- 	      & ---		  & ---\\
2010-Jul-01   &  56109.913 & 	---	      & 15.773$\pm$0.054  & --- 	      & ---		  & ---\\
2012-Jul-07   &  56115.892 & 18.863$\pm$0.088 & 15.777$\pm$0.056  & --- 	      & ---		  & ---\\
2012-Jul-10   &  56118.885 & 18.726$\pm$0.087 & 15.795$\pm$0.051  & --- 	      & ---		  & ---\\
2013-Mar-13   &  56364.225 &  --- 	      & ---               & 12.504$\pm$0.015  & 11.897$\pm$0.025  & 11.308$\pm$0.020 \\
2013-Apr-21   &  56403.164 & 18.582$\pm$0.042 & 15.666$\pm$0.026  & --- 	      & ---		  & --- \\
2013-Apr-22   &  56404.145 &   --- 	      & ---               & 12.591$\pm$0.007  & ---        	  & 11.278$\pm$0.060 \\
2013-May-03   &  56415.075 &   --- 	      & ---               & 12.446$\pm$0.057  & ---               & 11.202$\pm$0.084 \\
2013-Jun-30   &  56473.074 &   ---            & ---		  & 12.325$\pm$0.006  & 11.727$\pm$0.071  & 11.227$\pm$0.059 \\
2013-Jul-12   &  56485.011 &   --- 	      & ---               & 12.464$\pm$0.016  & ---		  & ---\\  
2013-Aug-11   &  56515.996 &   ---	      & ---		  & 12.493$\pm$0.022  & 11.262$\pm$0.068  & --- \\
2013-Aug-28   &  56532.892 &   ---	      & ---		  & 12.517$\pm$0.020  & 11.817$\pm$0.009  & 11.255$\pm$0.037 \\
2013-Sep-09   &  56544.937 &   ---	      & ---		  & 12.424$\pm$0.006  & 11.688$\pm$0.005  & 11.196$\pm$0.023  \\	
2013-Sep-11   &  56546.022 &   ---	      & ---		  & 12.324$\pm$0.049  & 11.745$\pm$0.009  & 11.209$\pm$0.054 \\       
2013-Sep-11   &  56546.942 &   ---	      & ---		  & 12.463$\pm$0.031  & 11.845$\pm$0.009  & 11.278$\pm$0.064 \\       
2013-Oct-09   &  56574.887 &   ---	      & ---		  & ---	 	      & 11.799$\pm$0.044  & 11.286$\pm$0.063  \\      
2013-Nov-01   &  56597.927 &   ---	      & ---		  & 12.493$\pm$0.035  & 11.797$\pm$0.007  & 11.269$\pm$0.095  \\	  
2013-Nov-10   &  56606.872 &   ---	      & ---		  & 12.455$\pm$0.070  & 11.826$\pm$0.090  & 11.294$\pm$0.007  \\      
2014-May-06   &  56783.146 & 18.582$\pm$0.047 & 15.683$\pm$0.051  & --- 	      & ---		  & ---\\	      
2014-Jun-04   &  56812.110 & 18.470$\pm$0.058 & 15.591$\pm$0.054  & --- 	      & ---		  & ---\\	           
2014-Aug-02   &  56871.111 & 18.416$\pm$0.043 & 15.565$\pm$0.051  & --- 	      & ---		  & ---\\
2014-Aug-15   &  56884.073 & 18.503$\pm$0.049 & 15.489$\pm$0.051  & --- 	      & ---		  & ---\\
2014-Sep-03   &  56903.022 & 18.378$\pm$0.044 & 15.537$\pm$0.051  & --- 	      & ---		  & ---\\
2014-Sep-03   &  56903.026 & 18.433$\pm$0.045 &   ---             & --- 	      & ---		  & ---\\
2014-Sep-23   &  56923.926 & 18.441$\pm$0.042 & 15.540$\pm$0.051  & --- 	      & ---		  & ---\\
2014-Oct-10   &  56940.007 & 18.272$\pm$0.094 & 15.501$\pm$0.054  & --- 	      & ---		  & ---\\
2014-Nov-17   &  56978.825 & 18.558$\pm$0.042 & 15.646$\pm$0.051  & ---		      & ---		  & ---\\

\hline
\end{tabular}
\label{optical_mag}
\end{center}
\end{table*}

\section{Observations and  data reduction}\label{obs}

\subsection{Optical spectroscopic observations}  \label{specobs}
 
Optical spectroscopic observations of the companion were performed during 2011 April -- 2014 November with four 
different telescopes: the Russian-Turkish 1.5-m telescope (RTT150) at the 
T\"{U}B\.{I}TAK National Observatory in Antalya (Turkey), the 2.56-m  Nordic Optical Telescope (NOT) located at 
the Observatorio del Roque de los Muchachos (La Palma, Spain), and the the 1.5-m Telescope at the Observatorio de 
Sierra Nevada (OSN-CSIC) in Granada (Spain). In addition to this long-term observation set we include a spectrum of the 
source taken in June 2007 with RTT150. 

The spectroscopic data from  RTT150 were obtained with the T\"{U}B\.{I}TAK Faint Object Spectrometer
and Camera (TFOSC). It is equipped with a 2048$\times$2048, 15$\mu$m pixel Fairchild 447BI CCD whose FOV is 
13$\farcm$3 $\times$ 13$\farcm$3. We used slit 67 $\mu$m (1$\farcs$24) with Grism 8 having an average dispersion 
of 1.1 \AA~pixel and providing a 5800--8300 \AA~wavelength coverage. The reduction of RTT150 spectra was done using 
the Long-Slit package of MIDAS.\footnote{http://www.eso.org/projects/esomidas}. Bias correction, flat-fielding and removal 
of cosmic-ray hits were carried out with standard MIDAS routines. 

The low-resolution OSN spectra (R$\approx$1400) were acquired using Albireo spectrograph 
centred on H$\alpha$ wavelength (6562.8 \AA) whereas NOT spectrum was obtained with the Andaluc\'ia Faint Object 
Spectrograph and Camera (ALFOSC),\footnote{The data presented here were obtained [in part] with ALFOSC, which is provided by 
the Instituto de Astrof\'isica de Andaluc\'ia (IAA) under a joint agreement with the University of 
Copenhagen and NOTSA.} using Grism\,7, with a dispersion of 1.5\,\AA/pixel, and 0$\farcs$5--1$\arcsec$ slits. The reduction of this data set was performed  using standard procedures within IRAF,\footnote{IRAF 
is distributed by the National Optical Astronomy Observatory, optical images which is operated by the 
Association of Universities for Research in Astronomy (AURA) under cooperative agreement with the National 
Science Foundation.} including bias subtraction, removal of pixel-to-pixel sensitivity variations, optimal 
spectral extraction, and wavelength calibration based on arc-lamp spectra. 

All spectroscopic data were normalized with a spline fit to continuum and corrected to the barycenter after
the wavelength calibration. The full width at half maximum (FWHM) and EW measurements
of H$\alpha$ lines were acquired by fitting Gaussian functions to the emission profiles using the ALICE 
subroutine of MIDAS.

\subsection{Optical/IR photometric observations}\label{photobs}
 As a part of our monitoring campaign the optical counterpart to \bex~ has been observed in the optical and  infrared bands 
during the period of 2011 September -- 2014 November (see Table~\ref{optical_mag}) with the 80-cm IAC80 and the 1.5-m TCS telescopes at the Observatorio del Teide on Tenerife (Spain) respectively. We obtained the optical photometric CCD images using B and V filters with integration time of 120\,s. In infrared, J, H and K$_{s}$ simultaneous observations were performed using the CAIN camera with integration  times of 150\,s. The reduction of the data was done by using the pipelines of both telescopes based on the 
standard aperture photometry \citep[for more details on reduction]{camero14}.  

The main part of the long-term optical CCD observations of the source include the 0.45-m reflecting 
ROTSEIIId\footnote{The Robotic Optical Transient Search Experiment, ROTSE, is a collaboration of 
Lawrence Livermore National Lab, Los Alamos National Lab, and the University of Michigan 
(http://www.ROTSE.net)} telescope data achieved from 2005 April to 2012 November (MJD 53465--56215). 
ROTSEIIId telescope, located at the T\"{U}B\.{I}TAK National Observatory (Antalya, Turkey), operates 
without filters is equipped with a 2048 $\times$ 2048 pixel CCD. For a total field of view (FOV) 
1$\fdg$85 $\times$ 1$\fdg$85, the pixel scale is defined as 3$\farcs$3 pixel$^{-1}$ \citep{akerlof03}.
During the observations, a total of 2014 CCD frames were collected covering almost 8 seasons. Dark and 
flat-field corrections of all images were done automatically by a pipeline immediately after the pointing.  
Instrumental magnitudes of all the corrected images were obtained using an aperture of 3 pixels (10 arcsec) in 
diameter by SExtractor Package \citep{bertin96}. By comparing all the stars in each frames with 
USNO-A2.0 catalog $R$--band magnitudes, calibrated ROTSEIIId magnitudes were acquired. 
For the timing analysis the time series were corrected to the barycenter by using JPL DE200 ephemerides 
\citep[for the details of ROTSEIIId data reduction]{kiziloglu05}.

\subsection{X-ray observations}\label{xobs}

The $Chandra$ X-ray Observatory observed XTE J1946$+$274 with the Advanced CCD 
Imaging Spectrometer (ACIS) instrument in FAINT mode in 2013 March 12 (MJD 56363.115) for a total exposure 
time of 4.6\,ks. Figure~\ref{plot_log} and Table~\ref{log_chandra} provide the log of this observation. 
We used a 1/8 subarray, which provides a time resolution of 0.4 s, and the typical ACIS-S imaging and 
spectral configurations. The source was positioned in the back-illuminated ACIS-S3 CCD at the nominal 
target position. Standard processing of the data was performed by the \textit{Chandra} X-ray Center 
(CXC) to Level 1 and Level 2 (processing software DS ver. 8.5.1.1). In this work we have used 
CIAO software (ver.\,4.6) for the reprocessing and the analysis of the data.
Since 2008, the Gamma-ray Burst Monitor (GBM) on board the \textit{Fermi} satellite, 
has been monitoring \bex. In this study we used timing products provided by the GBM Pulsar Team 
\citep[see e.g.][for a detailed description of the timing technique]{finger09,camero102}. 
We also used quick-look X-ray results provided by the \textit{RXTE} All Sky Monitor 
team,\footnote{http://xte.mit.edu/ASM$\_$lc.html}  and \textit{Swift}/BAT transient monitor 
results provided by the \textit{Swift}/BAT team \citep{krimm13}.

\begin{figure}
\hspace{-0.3cm}\includegraphics[width=9.5cm,height=12cm]{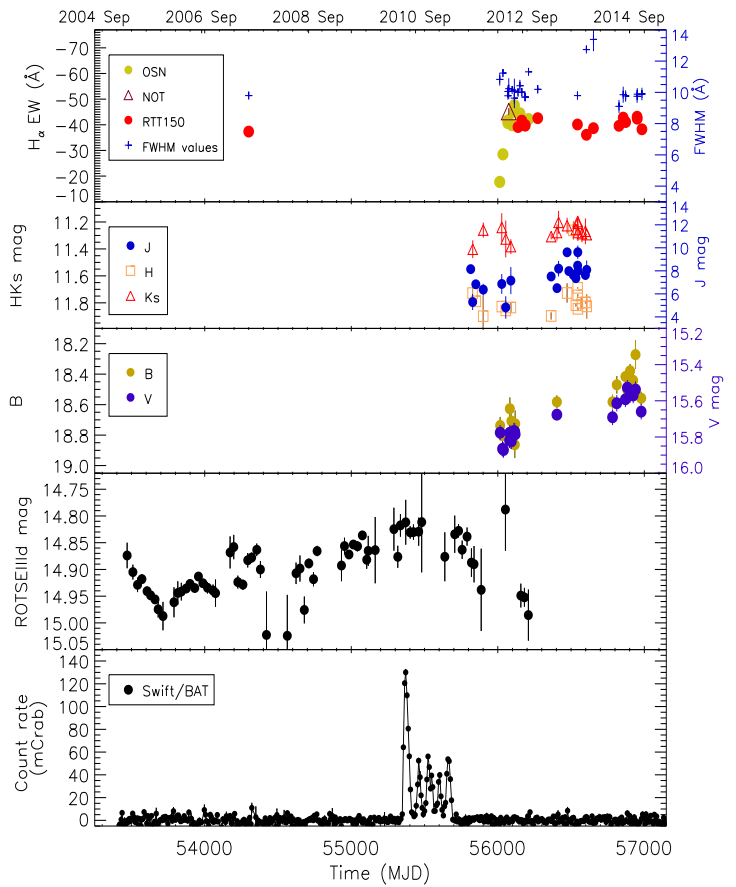}
\caption{Optical/IR and X-ray evolution of XTE\,J1946+274. The time bin for the ROTSEIIId lightcurve is 30 days and for \textit{Swift}/BAT is 8 days.} ¨
\label{opt_ir}
\end{figure}

\begin{figure}
\includegraphics[width=9.5cm,height=7cm]{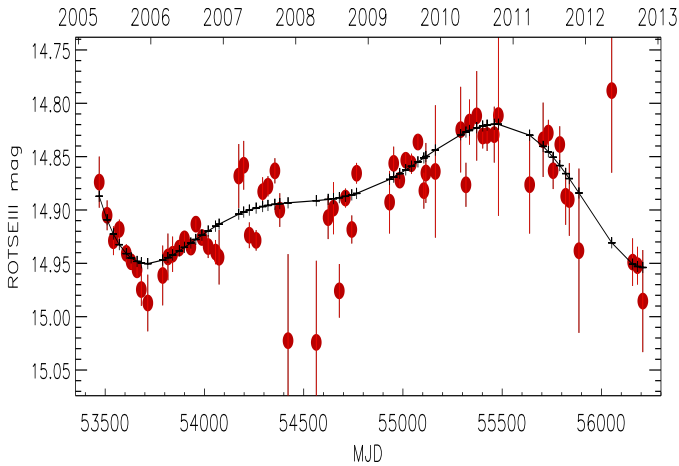}
\caption{Complete ROTSEIIId light curve binned using 30 days bins. Overplotted is the fitted sinusoid to the data (see Sect.~\ref{ana} for more details).}\label{period04}
\end{figure}


\begin{figure*}
\begin{center}
\begin{tabular}{lrr}
\includegraphics[scale=0.45,angle=-90,clip=true,trim=2mm 12mm 1mm 150mm]{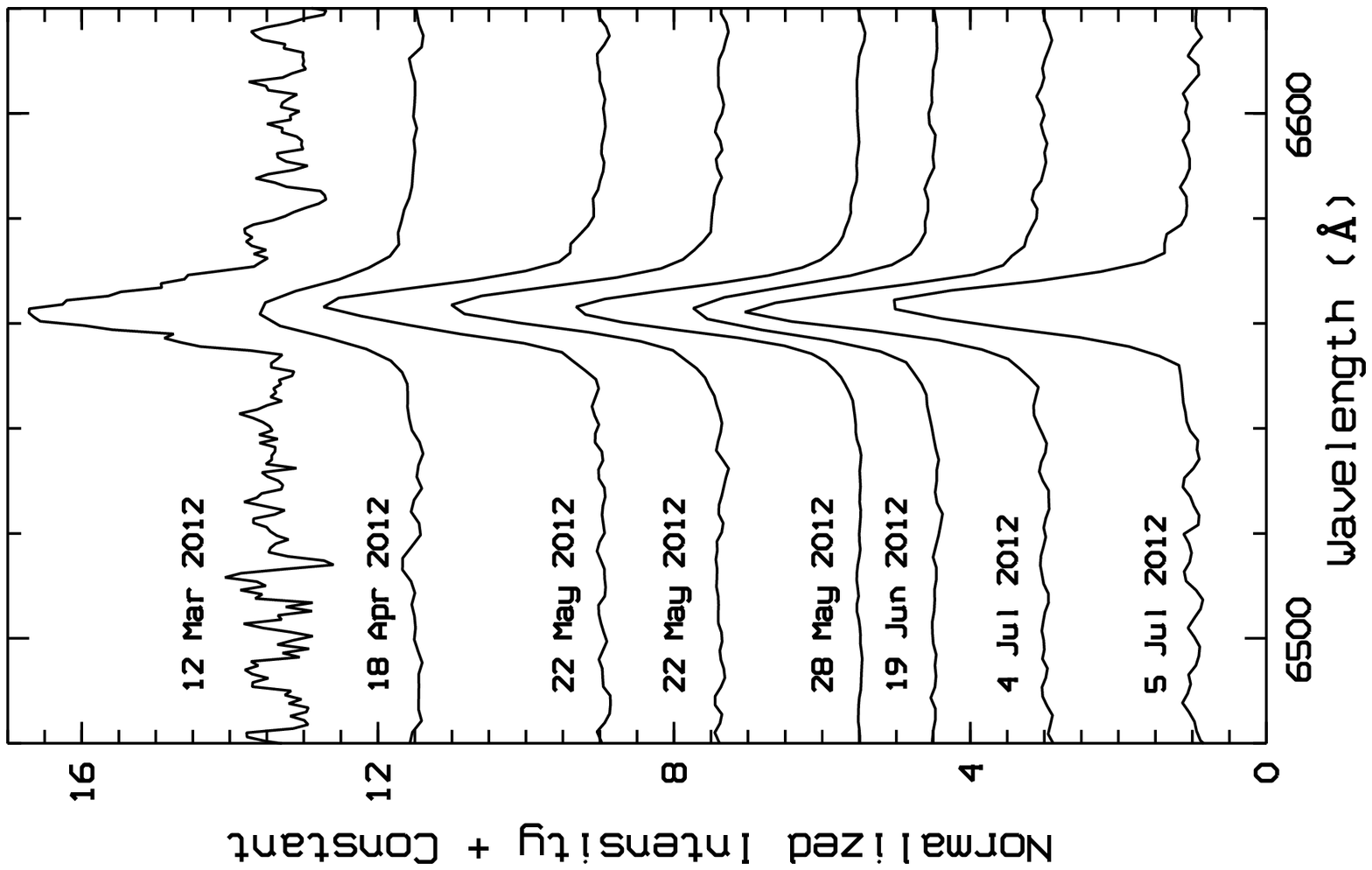} & \includegraphics[scale=0.45,angle=-90,clip=true,trim=2mm 12mm 1mm 150mm]{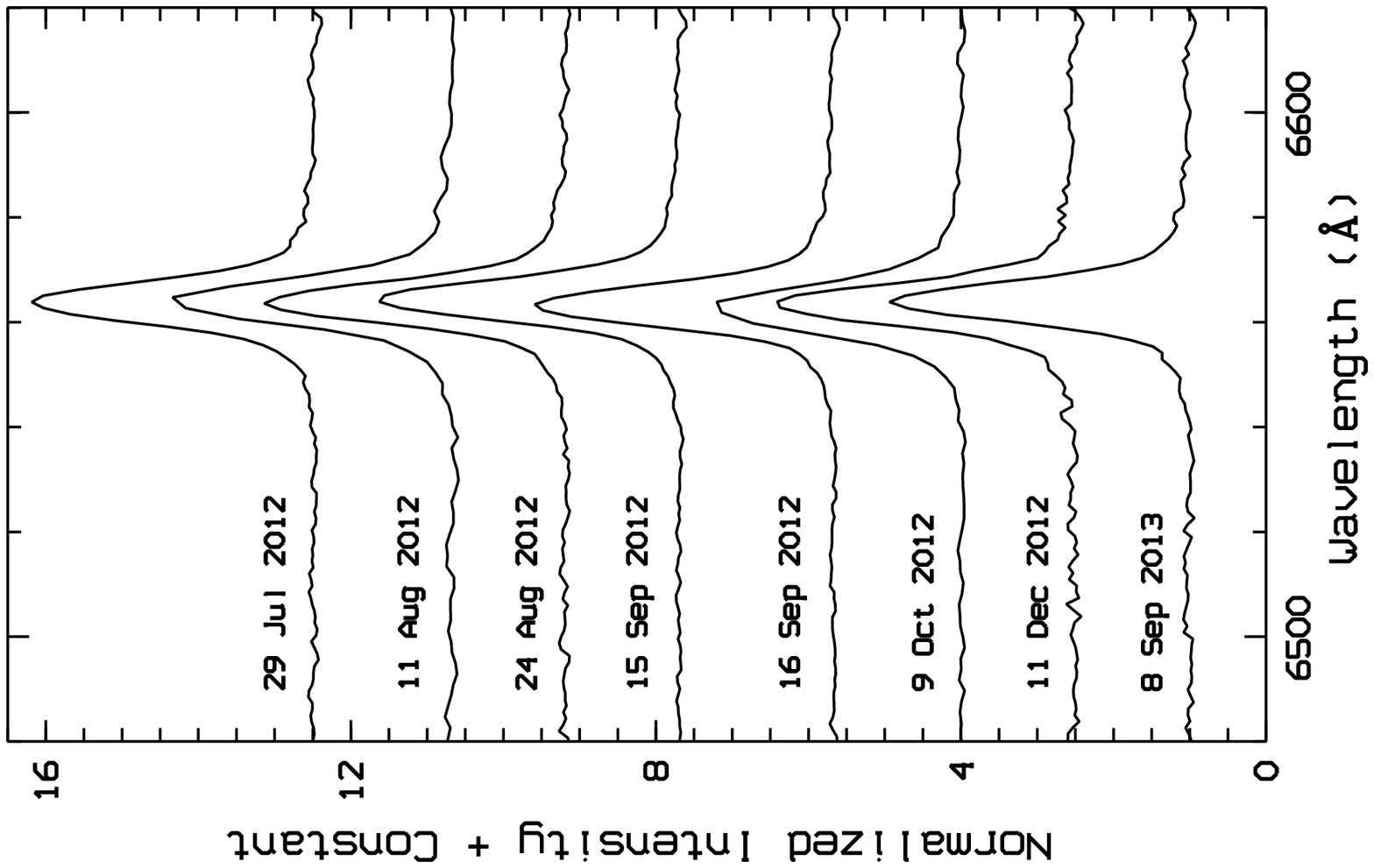}&\includegraphics[scale=0.45,angle=-90,clip=true,trim=2mm 12mm 1mm 150mm]{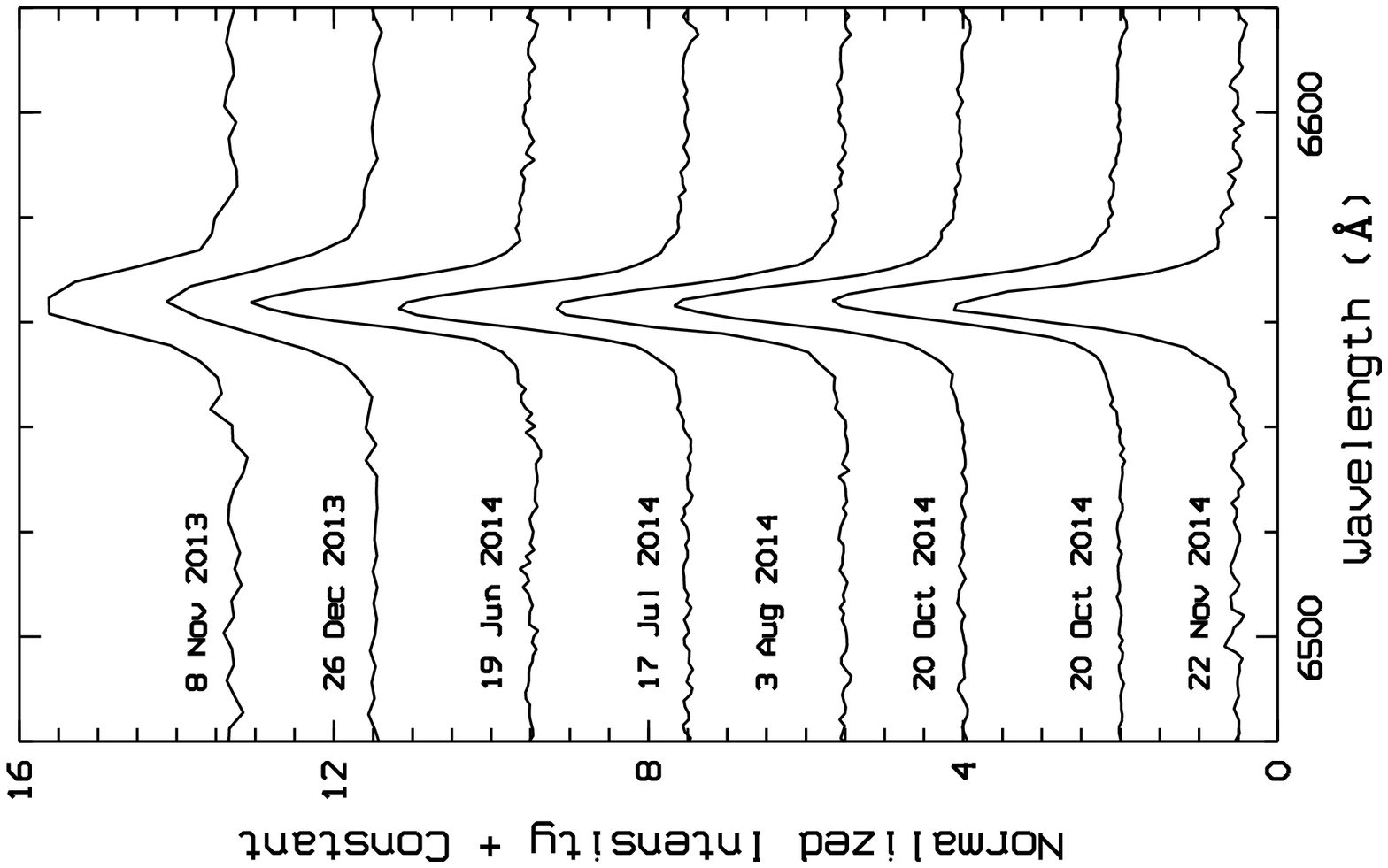}\\
\end{tabular}
\caption{H${\alpha}$ line profile evolution during the X-ray quiescence. The presence of any remarkable variation is not detected. }\label{H_quie2}
\end{center}
\end{figure*}

\section{Analysis and results}\label{ana}

\subsection{Optical/IR  photometry}\label{OIR}

The results of our multiwavelength campaign are shown in Fig.~\ref{opt_ir}. In this figure the optical/IR behavior of 
the Be star is displayed, together with the X-ray activity observed from the NS (see Tab.~\ref{optical_mag}). 
The ROTSE magnitude (fourth panel, from top to bottom) steadily varied from 2005 to 2013, with a minimum in 2006 and a 
maximum around mid 2010. A main X-ray outburst was then detected in 2010 lasting $\sim$60 days and reaching a flux 
of $\sim$140 mCrab in the 15--50 keV energy range (see bottom panel of Fig~\ref{opt_ir}). The following four outbursts 
were seen in a series, typical to Type I outbursts, with a separation of $\sim$60--90 days. In contrast to the other 
BeXRB systems showing recurrent series of normal outbursts, they did not coincide with the time of periastron passage 
of the NS. After the 2010/2011 X-ray outburst series, the Be/X-ray binary XTE J1946$+$274 did not show any transient 
activity. The quiescent phase has been ongoing for $\sim$4 years, although the optical/IR magnitudes have been 
showing an increasing trend since 2012 (see second and third panels of the same Figure). 

In order to search for periodic variability in the daily ROTSE light curve we used the Lomb-Scargle \citep{scargle82} and 
Clean \citep{roberts87} algorithms. The frequency analysis was applied to all photometric data, 
including 1747 points, over a range from 0 to 17.49 d$^{-1}$ (Nyquist frequency). We did not find any periodic 
variation in the original data, except for the power at the frequency of 1 d$^{-1}$, as a result of the daily observing 
schedule. We then rebinned the light curve in 30 days bins. Figure~\ref{period04} reveals a quasi-sinusoidal pattern 
in the data, with a minimum at the beginning of 2006 (MJD $\sim$53700). This seems to indicate that a brightening phase 
of the Be star occurred for almost five years, followed by a quick decline in about 2 years. To clarify this point 
we used the Fourier analysis module included in the \texttt{Period04} software 
package\footnote{http://www.astro.univie.ac.at/dsn/dsn/Period04} \citep{Lenz05}, which is based on a 
discrete Fourier transform algorithm. We thus fitted a sinusoid to a combination of the observed 
frequency 3.3(8)$\times$10$^{-4}$ d$^{-1}$ ($\sim$3030\,d), found by \texttt{Period04}, together with its first 
and second harmonics. The uncertainties in the frequency determinations were computed from Monte Carlo simulations. 
The resulted fitted curve (see also Figure~\ref{period04}) shows the probable evolution of the Be star brightening 
episode, probably a mass ejection event.
 

\subsection{H${\alpha}$ line} \label{halpha}


In Figure~\ref{H_quie2}, we present the spectroscopic tracing of the  
H${\alpha}$ line profiles observed between 2012--2014. The 
H${\alpha}$ line was always seen in a single-peaked emission (as the one in 2007 July 18) whereas EW and FWHM
measurements showed significant variations (see top panel of Figure~\ref{opt_ir}). The first
spectroscopic observation of this period had the lowest value of
H${\alpha}$~emission line, $\sim$18 $\AA$, ever observed for XTE\,J1946$+$274.
The weakness of this emission comparing to the typical values of
XTE J1946$+$274 might be interpreted as the variations in the decretion
disk of the Be star. Assuming that the EW measurement of H${\alpha}$ line emission was also
related to the amount of the material in the emitting region of
the disk, the weakness of the EW would be the result of the mass
loss either through the accretion of the NS or the truncation of its
size despite the lack of an X-ray activity. It is also possible that
the weakest value we caught does not represent the lowest one,
instead it can be a part of the refilling process of the disk after
a disk-loss episode. In fact, a sharp increasing trend of the EW
right after this value confirms the suggested idea. The increase
in EW lasted about three months reaching its peak value of $\sim$48 $\AA$. It is also important to note that this is the 
highest EW value of XTE J1946$+$274 ever observed. Although the line measurements
were scattered between MJD 56069--56209, they were not
significantly different than the average EW value.

During the observations of October and December 2013, the
widest emissions were detected while EW values were around
the average (see top panel of Figure~\ref{opt_ir}). In general the H${\alpha}$ line was seen as a narrow single-peaked
emission with an average FWHM value of $\sim$10.5 $\AA$.
It is also important to note that the EW and FWHM values of
the source show an inverse relation despite the expected positive
relation \citep{hanuschik89}.

In addition, we computed the rotational velocity of the Be star. First 
we estimated the projected rotational velocity of XTE J1946$+$274 as $vsini\sim$323 km s$^{-1}$ via the average 
values of EW and FWHM parameters \citep{hanuschik89}. Then using the inclination angle of the system given by Wilson 
t al. (2003), the true rotational velocity, $v_{rot}$, of the Be star was measured as 323--449 km\,s$^{-1}$. 
Taking the mass of the star as 16 M$_{\odot}$  and 8 R$_{\odot}$\,as the limit radius for a B type star, then 
the critical break-up velocity, $v_{crit}$, was found to be $\sim$618 km\,s$^{-1}$. Thus, we find the critical 
fraction, defined as the the ratio of the equatorial rotational velocity to the the break-up velocity, of XTE J1946$+$274 
as $w\sim$0.5--0.72. This result indicates that the Be star in XTE J1946$+$274 is rotating at 50--70$\%$ of its break-up 
velocity, typical to the stars for the same type.

\section{\textit{Chandra}/ACIS-S  X-ray analysis during quiescence}

\subsection{Imaging}

We extracted an image in the 0.3--10 keV energy range, using the \textit{Chandra} 
pointing observation from 2013 as described in Sect.~\ref{xobs}. We then applied the CIAO \texttt{celldetect} tool 
to the $\sim$4.6\,ks ACIS-S cleaned image and found \bex~ at $\alpha =19^{\rm h}45^{\rm m}39^{\rm s}.34$ and
 $\delta =27^\circ21\arcmin55\arcsec.36$ (J2000) with a signal-to-noise ratio of 11.025, and a statistical 
error of 0.03 arcsec radius. In addition, another X-ray bright star was detected 
at $\alpha =19^{\rm h}45^{\rm m}34^{\rm s}.91$ and  $\delta =27^\circ18\arcmin18\arcsec.04$ with a signal-to-noise 
ratio of 5.2, and a statistical error of 0.1 arcsec radius.
 
Furthermore, XTE J1946$+$274 is the 2MASS\,19453935+2721555 star with a catalog position 
of $\alpha =19^{\rm h}45^{\rm m}39^{\rm s}.36$ and  $\delta =27^\circ21\arcmin55\arcsec.52$ (J2000) 
(statistical error of 0.06 arcsec radius) and the second star is the 2MASS\,19453491+2718183 with 
a catalog position of $\alpha =19^{\rm h}45^{\rm m}34^{\rm s}.92$ and  $\delta =27^\circ18\arcmin18\arcsec.30$ 
(0.06 arcsec error radius). We then performed a boresight correction of the field to refine \bex~ position 
and its error circle. Assuming a physical association between the 2MASS stars and the X-ray sources, the final position of
 \bex~ is $\alpha =19^{\rm h}45^{\rm m}39^{\rm s}.4$ and  $\delta =27^\circ21\arcmin55\arcsec.5$ with a 
1$\sigma$ associated error circle of 0.2 arcsec radius (computed doing a quadratic mean of all the 
positional and statistical errors plus the 2MASS catalogue intrinsic systematic errors).

\subsection{Spectral study}\label{spec}

To  obtain the 0.3--10\,keV phase-averaged spectrum for the \textit{Chandra} ACIS-S observation we used source and 
background photons extracted as described in Section~\ref{obs}. We used the 
\texttt{specextract} script, which uses a combination of CIAO tools, to extract source 
and background spectra for \bex. To extract only the photons from the point source  a circular region with 2$\arcsec$.5 
radius and a circular background region of radii 18$\arcsec$ (far from the source) were used. For the present analysis we 
used the {\tt XSPEC} package (version 12.8.1g) \citep{arnaud96}. 

Two models provided the best fit to the data, an absorbed power law (PL) and blackbody (BB) models. 
For the photoelectric absorption we used the cross-sections from \cite{balucinska92} and the Solar abundance 
from \cite{anders_grevesse89}. The best-fit parameters for these models can be seen in Tab.~\ref{spec_tab}. The main parameters for both models are $\Gamma$=0.9$\pm$0.4 (C-stat=108.32 for 148 d.o.f.), and  kT=1.43$\pm$0.17 keV (C-stat=109.15 for 148 d.o.f.), with $N_{\rm H}\sim$2--7$\times10^{22}$ cm$^{-2}$. The 2--10 keV observed flux was F$_X$=0.75(3)--0.97(2)$\times$10$^{-12}$\,erg cm$^{-2}$ s$^{-1}$ for 
the BB and PL models, respectively. The photon index is similar to the value
found with \textit{Swift}/XRT \citep{muller12} in the 1.5--7 keV
energy range ($\Gamma$=0.84(8)), 
and \textit{Suzaku}-XIS in the 0.3--10 keV band
($\Gamma$=1.09(5)). However, the X-ray flux level of XTE J1946$+$274
in those observations was about 2 order of magnitude larger
($\sim$2$\times$10$^{-10}$\,erg cm$^{-2}$ s$^{-1}$).


\begin{table}
\begin{center}
\caption{Spectral parameters from an absorbed PL model and absorbed BB. Errors are given at the 90$\%$ confidence level.}                          
\begin{tabular}{lll} 
\hline\hline 
\hspace{-0.01cm}  Parameter    		&   PL       &  BB \\  
\hline\hline 
 \hspace{-0.01cm}$N^a_{\rm H}$  &    0.7(4)					&	0.21(15)			   \\
 \hspace{-0.01cm}$\Gamma$        & 	0.9(4)					&-			 \\
 \hspace{-0.01cm}$\Gamma_{norm}^b$ &  	1.03(3)		&	-		  \\

 \hspace{-0.01cm}$kT$(keV) &  	-		&		1.43(17)	  \\
  \hspace{-0.01cm}$kT_{norm}$ &  		-	&		0.020(8)		  \\  

\hspace{-0.03cm} abs.Flux$^{c}$ &  1.0(3)			&		0.81(16)		  \\
\hspace{-0.03cm} unabs.Flux$^{c}$ &  1.2(4)			&	0.84(18)			  \\
\hspace{-0.03cm}C-stat (d.o.f.)  &  108.32(148)	&	109.15(148)		  \\     
\hline							    				
\end{tabular}

$^{a}$ $\times10^{22}$\,cm$^{-2}$.  \\
$^{b}$ $\times10^{-4}$ photons keV$^{-1}$ cm$^{-2}$ s$^{-1}$ at 1\,keV.  \\
$^{c}$ $\times10^{-12}$ erg cm$^{-2}$\,s$^{-1}$ in the 0.3--10\,keV  energy  band.\\

 \label{spec_tab}
\end{center}  
\end{table}

\begin{figure}
\begin{center}
\hspace{-0.3cm}\rotatebox{-90}{\includegraphics[width=3.5cm,height=7cm]{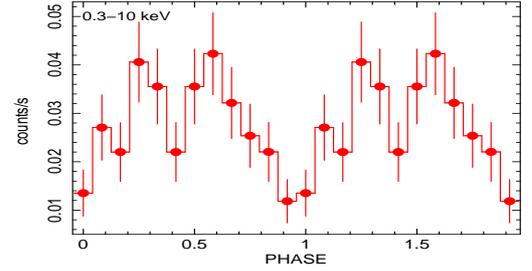}}
\caption{Background subtracted X-ray pulse profile (in counts s$^{-1}$) for XTE J1946$+$274 during quiescence in the 0.3--10 keV energy band.}\label{prof0}
\end{center}
\end{figure}

\begin{figure}
\begin{center}
\hspace{-0cm}\rotatebox{-90}{\includegraphics[width=8cm,height=6.7cm]{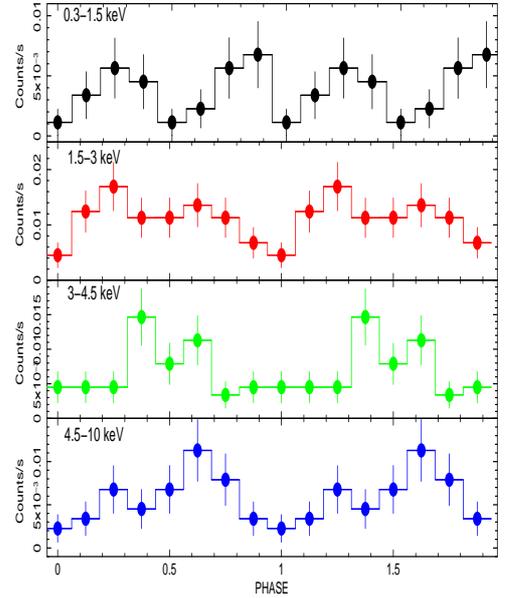}}
\caption{Background subtracted X-ray pulse profiles (in counts/s) during quiescence in different energy bands.}\label{prof_en}
\end{center}
\end{figure}

\begin{figure*}
\begin{center}
\includegraphics[width=18cm,height=9cm]{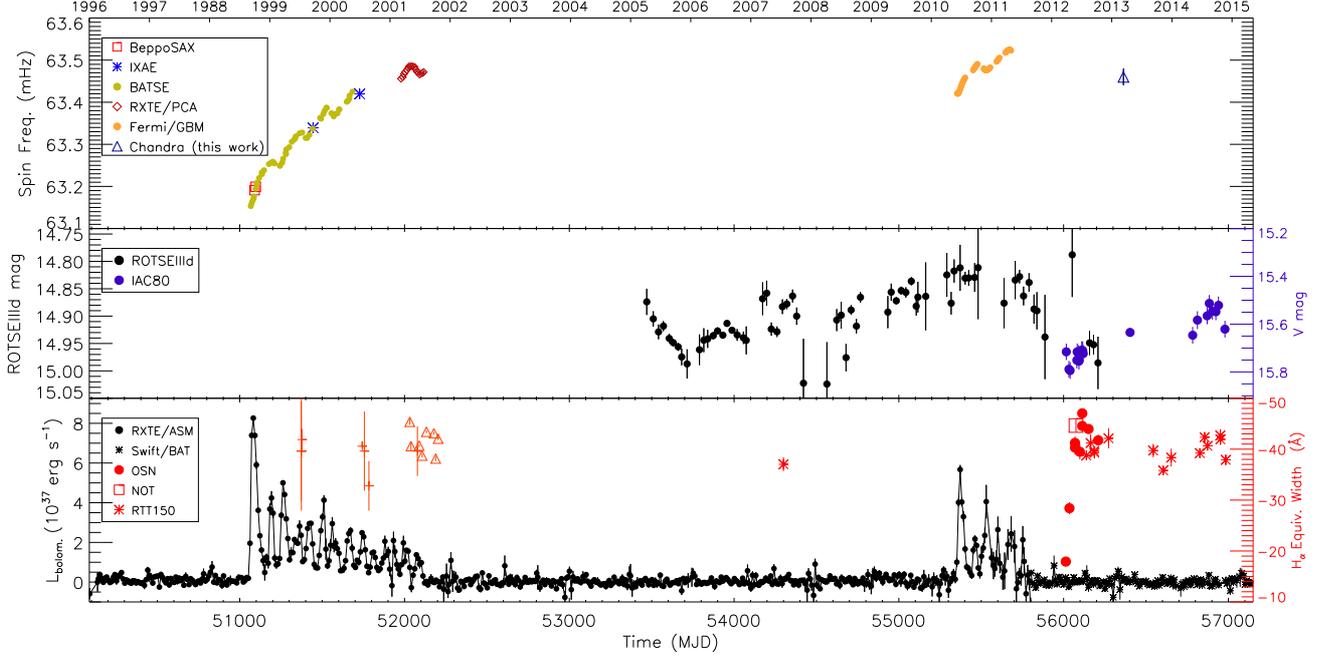}
\caption{\textit{Top}. Spin frequency history of XTE J1946$+$274 since its discovery in 1988. \textit{Middle}. Long-term optical light curve of this source. \textit{Bottom}. 
Long-term X-ray bolometric luminosity, and overplotted the evolution of the EW of the H$\alpha$ line (red stars, open square, and filled circles). 
The H${\alpha}$ EW measurements from 1999 to 2002  were extracted from \cite[][red crosses]{verrecchia02} and Wilson et al. 
(2003, red triangles). To compute the X-ray bolometric luminosity we follow the procedure by Wilson et al. (2003).}
\label{long_term}
\end{center}
\end{figure*}

\subsection{Timing}\label{timing}

For the timing analysis, we first referred the arrival time of each photon to the barycenter of 
the solar system using the CIAO tool  \texttt{axbary}. Then, we used the \texttt{dmextract} tool to 
create background-subtracted lightcurves, using the time resolution of the data ($\sim$0.4\,s). For this, 
we extracted the source photons on each individual observation from a circular region with 2$\farcs$5 radius, 
and another one for the background, far from the source.  

We searched for pulsations using the \texttt{Xronos} package and found a 
P$_{pulse}$=15.757(1)\,s (epoch= MJD 56\,363.115; $\nu$=0.063464(4) Hz). Figure~\ref{prof0} shows the pulse profile 
obtained by folding the X-ray data set using this period. Despite the low luminosity level of the 
source in this observation, we can observe a profile with two peaks separated by a dip at phase $\sim$0.4, and another 
dip at phase 0.95. Wilson et al. (2003) found that at its lowest intensity level during the 2001 outburst 
the profile consisted of an asymmetric structured main peak near phase 0.2 in an \textit{RXTE}/PCA observation of 
XTE J1946$+$274. We note, however, that this profile was obtained in the 2--30 keV energy range. 
In our \textit{Chandra}/ACIS-S observation the dip in the pulse profile was more prominent at low energies in 
the 0.3--1.5keV band, vanishing as the energy increases (see Fig.~\ref{prof_en}), and with the pulse profile 
evolving to single-peaked. The shape of the profile at 3--4.5 keV was narrower and quite symmetric comparing to 
the asymmetric single peaks found at 1.5--3.5 keV and 4.5--10 keV, with all peaking at different phases of 
$\sim$0.2, $\sim$0.4, and $\sim$0.6 for each energy band.

The modulation amplitude of the 0.3--10\,keV pulse displayed in Fig.~\ref{prof0} (12 phase bins) can be measured 
using a pulse fraction defined as follows,
\begin{equation}PF_{rms}=\frac{1}{\bar{y}}
\sqrt{\frac{1}{n}\sum\limits_{i=1}^n ((y_{i}-\bar{y})^{2} -\sigma_{i}^2})\end{equation}  
        
where  $n$ is the number of phase bins per cycle, $y_{i}$ is the number of counts in the $i$th phase bin, $\sigma_i$ is the error on $y_{i}$ and $\bar{y}$ is the mean number of counts in the cycle. 
Applying this formula to our data we obtained a $PF_{rms}$=32.1(3)$\%$. Using the standard definition, 
\begin{equation} PF_{std}=\frac{F_{max}-F_{min}}{F_{max}+F_{min}}\end{equation}          
the value we obtained was larger, as expected, being 56(20)$\%$. In addition, we found $PF_{rms}$ 
estimates of 51.02(12)$\%$, 18.52(2)$\%$,  54.27(13)$\%$, and 37.23(15)$\%$  in the 0.3--1.5\,keV, 1.5--3\,keV, 
3--4.5\,keV, and 4.5--10\,keV energy bands, respectively.  
We would like to note that for the profiles at different energy bands the phase binning was reduced to 8, since the $PF_{rms}$ formulation did not yield real solutions in some of the bands. 

\section{Discussion}

\subsection{Be/NS interaction}

XTE J1946$+$274 is one of the BeXRBs that spends most of its
time in an X-ray quiescent phase. The uniqueness of the system
comes from its X-ray outburst behavior that is not connected
to the orbital passages of the NS. The shifts in the
outburst phases with respect to the periastron/apastron passages
of the NS is thought to be result of the global perturbations triggered
by the truncation of the disk radius.  During quiescent state we have
not seen any trace of such density wave in the decretion
disk that shows itself as the variation of the emission profiles in
the spectroscopic data. Wilson et al. (2003) attributed the variations
in the H${\alpha}$ emission profile to the existence of the perturbations
in the disk and added the difficulties of detection of
the density perturbations in XTE J1946$+$274 since the relatively
small viewing angle restricts the size of the projected area to
be observed. According to that, only the large-scaled perturbations
can be seen in the disk of XTE J1946$+$274. In contrast to
this idea, \cite{silaj10} suggested that emission line profile
shapes could not be used to estimate the inclination angle of the
system since for a given inclination angle different types of profile
shapes might be produced as a result of the density changes
in the disk-thermal structure. This means that, in the
suggested picture of Wilson et al. (2003) for XTE J1946$+$274,
we should have seen the profile changes if there had occurred any
density variations in the decretion disk, despite the small inclination
angle of the system. Therefore it is very likely not to see any profile changes in the emission lines. 

Be stars are variable on all time scales. Short-term variability is commonly associated with either the rotation 
of the star and/or non-radial pulsations \citep[see e.g.][]{kiziloglu07,juanguso11}, while the long-term variability is believe 
to be originated by structural changes in the Be decretion disk. The typical time scale for the formation and disintegration of 
the disk has been found to coincide with the observed time scales of the largest amplitude variations \citep{lyuty00,camero14}. 
The observed long optical brightening experienced by XTE J1946$+$274 may be interpreted as a slow mass ejection event, 
which started in 2006, peaked in 2010 and rapidly decreased afterwards, reaching a quiescence level in 2012. 

The high values obtained for the H${\alpha}$ EW, with the line always seen in emission, may indicate 
that a large Be circumstellar disk is permanently present in this system. It is also worth noting that although 
renewed X-ray activity was 
not observed until 2010, the EW value of the H${\alpha}$ line was already --37.35 $\AA$ in 2007 July 
(see Table~\ref{xte_ew}). It is possible that the decretion disk of the Be star in XTE J1946$+274$ is tilted with 
respect to the orbital motion of the 
NS. In such a misaligned system warping of the differentially rotating disk can trigger the X-ray activity when 
the NS capture material 
from this area \citep[see][]{okazaki13}. The 2010 X-ray outbursts series 
emptied accumulated material in the warped regions and the disk dramatically decreased around 2011 (see bottom panel of  Fig~\ref{long_term}). Although the 
recovery of the decretion disk was fast, our data point out that the disk reached its maximum size by the time the mass ejection 
event was over. In such a stable environment the disk did not get warped and therefore no X-rays were detected.
This seems to be confirmed by the absence of any variations in our H${\alpha}$ line profiles during the same period. 

The ongoing optical/IR brightening, probably initiated around mid 2012, resembles so far the 
previous mass ejection event between 2006--2012. If the phenomenon repeats we foresee that approximately 
in two years XTE J1946$+$274 will manifest in X-rays.

\subsection{NS behavior}

XTE J1946$+$274 displayed an extended period of activity from 1998 September to 2001 July. Wilson et al. (2003) 
noted that this period of activity resembled a series of normal outbursts (thirteen) more than a single giant outburst.
After approximately 9 years in quiescence, XTE J1946$+$274 reawakened in X-rays with a series of five outbursts between 2010 
and 2011 (but less intense than in 1998). Once again, two outbursts were observed per orbital period, which did not clearly 
coincide with the times of periastron and apastron passages of the NS. \textit{Fermi}/GBM detected pulsations from XTE 
J1946$+$274 between 2010 and 2011 (see top panel of Fig.~\ref{long_term}). 

The last frequency measurement reported by GBM was 0.0635228(3) Hz (P$_{spin}$=15.74238(2)\,s) in 2011. 
Although the frequency histories were not orbitally corrected, and therefore a strong orbital modulation 
is contaminating the intrinsic torque variations, considering our frequency determination of 2013, the source might have spun down at a rate
close to $\dot{\nu}\sim-0.98\times$10$^{-11}$ Hz\,s$^{-1}$ ($\dot{P}\sim2.4\times$10$^{-10}$ s\,s$^{-1}$). Spin-up torque/flux correlations for this source were found by
\cite{wilson03}, suggesting the presence of an accretion disk around the NS during X-ray active periods. This scenario is corroborated by \textit{Fermi}/GBM 
during the 2010/2011 X-ray outbursts (see Fig.~\ref{long_term}).

To determine whether the observed X-ray turn-off was due to centrifugal inhibition of accretion or not \citep{stella86}, 
\cite{wilson03} estimated the flux at the onset of this effect by equating the magnetospheric radius and the corotation 
radius, obtaining a threshold flux for the onset of the centrifugal inhibition of accretion 
in the range of 0.6--6.0$\times$10$^{-11}$ erg cm$^{-2}$ s$^{-1}$. Our observed flux during 
quiescence in 2013 was well below this range ($\sim 0.1\times$10$^{-11}$ erg cm$^{-2}$ s$^{-1}$), corresponding to 
an X-ray luminosity of $\sim$1.2$\times$10$^{34}$ erg s$^{-1}$ in the 0.3--10 keV range. 
We would like to point out that the transition flux level depends on several factors 
\citep[see equation 17 in][]{wilson03}, although the authors only considered the uncertainty of one of them 
(the constant factor k; see also equation 9 in Wilson et al., 2003). To compute a good estimate of the 
transition flux error it is not an easy task, mainly because there is a considerable uncertainty 
in the NS parameters. For instance, recent NS mass measurements in X-ray/optical binaries give a range of 
M=[1.037,2.44]M$_{\odot}$ (mean value 1.568 M$_{\odot}$ ; \cite{latti12}). In addition, \cite{cipo15} equation 
of state (EOS) predicts a range of values for the NS radius of  7.64--18.86 km. In particular, \cite{sharma15}
found in their model that for a NS of 1.5 M$_{\odot}$ the predicted radius should be of the order of 11.67 km. Having this in 
mind and assuming an error on the NS parameters of about 20 $\%$, we would obtain a lower limit for the transitional flux 
$\sim$0.12$\times10^{-11} erg^{-2} s^{-1}$, in agreement with the quiescent flux observed by {\it Chandra} considering 
the uncertainties.  The flux lower limit would drop with higher errors on the NS parameters. Therefore, 
the compact object on XTE J1946$+$274 might have not been in centrifugal inhibition of accretion in 2013 but close to 
enter in that regimen.

Based on our observation of XTE J1946$+$274 during quiescence in 2013, and 
assuming that the source enters the supersonic propeller regime \citep[see e.g.][]{davies_pringle81, henrichs83}, 
it is expected to spin down at a rate of $\dot\nu_{super}=(-4\pi\nu^2\mu^2\,(GM)^{-1}I^{-1})(c_s/\mu r_A)$, where $\mu$ is the NS magnetic moment, $M$ the mass (1.4 M$_{\odot}$), $I$ the moment of inertia, $c_{s}$ is the sound speed at the magnetospheric radius (which we take here of the order of the free-fall velocity), and r$_A$ is the Alfv\'{e}n radius \citep{henrichs83}. 
With this $\dot\nu_{super}$ found to be $\thickapprox\ -0.13\times$10$^{-12}$ Hz\,s$^{-1}$. 
We note, however, that there is no general consensus on the estimate of the 
propeller efficiency, and the difference on the estimate of the torque
during supersonic propeller state may be as large as $\thickapprox$10$^{4}$ under
certain conditions \citep[see discussion for LS I+61$^{\circ}$303 in][]{papitto12}. It is also possible that other types of torques might be operating simultaneously with the propeller-type torque, e.g. the
"frictional" torque \citep[][and references therein]{ghosh95}, with
the effect of increasing the observed spin-down torque on the NS. On the other hand, the X-ray luminosity  in supersonic propeller regimen is given by L$_X$=8\,$\dot{M}\,v^{2}_{rel}$ \citep{henrichs82}, where $\dot{M}$ is the mass accretion rate, and $v_{rel}$ is the relative velocity which is typically of the order of $\sim$1000\,km/s. 
For XTE\,J1946$+$274 this yields to L$_{X}\thickapprox$1.13$\times$10$^{34}$ erg\,s$^{-1}$, which is close to what was observed
in 2013 ($\sim$1.19$\times$10$^{34}$ erg\,s$^{-1}$). Moreover, we found that this is
also true for the long quiescence period between 2001 and 2010.
The observed spin-down rate from MJD 55353.01 to 52115.70,
computed using the last frequency determination by Wilson et al. (2003) in 2001 ($\sim$0.0635 Hz) and the first one detected by GBM ($\sim$0.063421 Hz) in 2010, gives $\dot\nu_{obs}\thickapprox\,0.18\times$10$^{-12}$ Hz\,s$^{-1}$. 
This result is also close to the observed spin-down rate during this particular period.

\section{Summary and conclusions}

We have performed a long-term multiwavelength study of the Be/X-ray binary system XTE J1946+274, a source which spends almost all its time in quiescence. Our findings include the detection of a large Be circumstellar disk during that phase. We address the absence of major X-ray activity by discussing our results in terms of the neutron star Be-disk interaction. 

$\bullet$ A long mass ejection event from the Be-star commenced in 2006, attained its maximum intensity in 2010, and reached 
quiescence presumably in 2012.  

$\bullet$ The high values obtained for the H${\alpha}$ EW (always in emission) point to the permanently presence of a large 
Be circumstellar disk, probably tilted with respect to the orbital motion of the NS. 


$\bullet$ We proposed that the mechanism that might be triggering the X-ray activity is 
the contact of the NS with the warped area of the Be disk. After the series of X-ray episodes, 
the warped regions emptied and the disk dramatically decreased.  

$\bullet$ There was an absence of variations in the H${\alpha}$ line profiles during the posterior fast recovery of the disk. This took place in parallel to the decay of the optical mass ejection event. In a much more stable environment the disk did not get warped and therefore no X-rays were detected.  

$\bullet$ On the other hand, pulsations from the NS have been detected during X-ray quiescence  in 2013, with P$_{pulse}$=15.757(1)\,s and an rms pulse fraction of 32.1(3)$\%$. 

$\bullet$ The 0.3--10 keV X-ray spectrum of the NS was well fitted with either an absorbed  black-body or absorbed power-law models (kT=1.43$\pm$0.17 and $\Gamma$=0.9$\pm$0.4).  

$\bullet$ The observed X-ray luminosity during quiescence is close to that  observed  when the NS is in supersonic propeller regimen.

$\bullet$ The ongoing optical/IR brightening resembles so far the previous mass ejection event. If the proposed scenario from above is valid we predict that approximately in two years XTE J1946$+$274 will show major X-ray activity. \\\\

Acknowledgments. We thank the anonymous referee for his/her helpful comments and suggestions. 
This article is partially based on service observations made with the IAC80 and TCS telescopes operated on the island of Tenerife by the Instituto de Astrof\'{\i}sica de Canarias (IAC) in the Spanish Observatorio del Teide. The present work is also based on observations made with the Nordic Optical Telescope, operated by the Nordic Optical Telescope Scientific Association at the Observatorio del Roque de los Muchachos (IAC), La Palma, Spain. The Albireo spectrograph at the 1.5-m telescope is operated by the Instituto de  Astrof\'{\i}sica de Andaluc\'{\i}a at the Sierra Nevada Observatory.
 We thank T\"{U}B\.{I}TAK and ROTSE collaboration for partial support in using the RTT 150 and ROTSEIIId
 Telescopes with project numbers TUG-RTT150.08.45, 12ARTT150-264-1 and ROTSE-40. 
The scientific results reported in this article are based in part on data obtained from 
the \textit{Chandra Data Archive}. The work of J.G.S. is supported by the Spanish Programa Nacional de 
Astronom\'{\i}a y Astrof\'{\i}sica under contract AYA2012-39246-C02-01. E. N. acknowledges a VALi+d 
postdoctoral grant from the Generalitat Valenciana and was supported by the Spanish Ministry of Economy 
and Competitiveness under contract AYA 2010-18352. A. C. was supported by the AYA2012-39303, SGR2009-811 
and iLINK2011-0303 grants. M.\"{O}. A. acknowledges support from T\"{U}B\.{I}TAK, The Scientific and 
Technological Research Council of Turkey, through the research project 106T040. 


\begin{thebibliography}{40}
\expandafter\ifx\csname natexlab\endcsname\relax\def\natexlab#1{#1}\fi

\bibitem[{{Akerlof} {et~al.}(2003){Akerlof}, {Kehoe}, {McKay}, {Rykoff},
  {Smith}, {Casperson}, {McGowan}, {Vestrand}, {Wozniak}, {Wren}, {Ashley},
  {Phillips}, {Marshall}, {Epps}, \& {Schier}}]{akerlof03}
{Akerlof}, C.~W., {Kehoe}, R.~L., {McKay}, T.~A., {et~al.} 2003, \pasp, 115,
  132

\bibitem[{{Anders} \& {Grevesse}(1989)}]{anders_grevesse89}
{Anders}, E. \& {Grevesse}, N. 1989, \gca, 53, 197

\bibitem[{{Arnaud}(1996)}]{arnaud96}
{Arnaud}, K.~A. 1996, in Astronomical Society of the Pacific Conference Series,
  Vol. 101, Astronomical Data Analysis Software and Systems V, ed. G.~H.
  {Jacoby} \& J.~{Barnes}, 17

\bibitem[{{Balucinska-Church} \& {McCammon}(1992)}]{balucinska92}
{Balucinska-Church}, M. \& {McCammon}, D. 1992, \apj, 400, 699

\bibitem[{{Bertin} \& {Arnouts}(1996)}]{bertin96}
{Bertin}, E. \& {Arnouts}, S. 1996, \aaps, 117, 393

\bibitem[{{Caballero} {et~al.}(2010){Caballero}, {Pottschmidt}, {Bozzo},
  {Ferrigno}, {Neronov}, {Santangelo}, {Klochkov}, {Staubert}, {Kretschmar},
  {Wilms}, {Kreykenbohm}, {Fuerst}, {Schoenherr}, {Rothschild}, \&
  {Suchy}}]{caballero10}
{Caballero}, I., {Pottschmidt}, K., {Bozzo}, E., {et~al.} 2010, The
  Astronomer's Telegram, 2692, 1

\bibitem[{{Camero} {et~al.}(2014){Camero}, {Zurita}, {Gutierrez Soto}, {Ozbey
  Arabaci}, {Nespoli}, {Kiaeerad}, {Beklen}, {Garcia-Rojas}, \&
  {Caballero-Garcia}}]{camero14}
{Camero}, A., {Zurita}, C., {Gutierrez Soto}, J., {et~al.} 2014, ArXiv e-prints

\bibitem[{{Camero-Arranz} {et~al.}(2010{\natexlab{a}}){Camero-Arranz},
  {Finger}, {Ikhsanov}, {Wilson-Hodge}, \& {Beklen}}]{camero102}
{Camero-Arranz}, A., {Finger}, M.~H., {Ikhsanov}, N.~R., {Wilson-Hodge}, C.~A.,
  \& {Beklen}, E. 2010{\natexlab{a}}, \apj, 708, 1500

\bibitem[{{Camero-Arranz} {et~al.}(2010{\natexlab{b}}){Camero-Arranz},
  {Finger}, {Wilson-Hodge}, \& {Jenke}}]{camero101}
{Camero-Arranz}, A., {Finger}, M.~H., {Wilson-Hodge}, C., \& {Jenke}, P.
  2010{\natexlab{b}}, The Astronomer's Telegram, 2677, 1

\bibitem[{{Campana} {et~al.}(1999){Campana}, {Israel}, \& {Stella}}]{campana99}
{Campana}, S., {Israel}, G., \& {Stella}, L. 1999, \aap, 352, L91

\bibitem[{{Cipolleta} {et~al.}(2015){Cipolletta}, {Cherubini}, {Filippi}, {Rueda}, {Ruffini}}]{cipo15}
{Cipolleta}, F., {Cherubini}, C., {Filippi}, S., {Rueda}, J.~A., {Ruffini}, R. 2015, arXiv:1506.05926

\bibitem[{{Davies} \& {Pringle}(1981)}]{davies_pringle81}
{Davies}, R.~E. \& {Pringle}, J.~E. 1981, \mnras, 196, 209

\bibitem[{{Finger} {et~al.}(2009){Finger}, {Beklen}, {Narayana Bhat},
  {Paciesas}, {Connaughton}, {Buckley}, {Camero-Arranz}, {Coe}, {Jenke},
  {Kanbach}, {Negueruela}, \& {Wilson-Hodge}}]{finger09}
{Finger}, M.~H., {Beklen}, E., {Narayana Bhat}, P., {et~al.} 2009, ArXiv
  e-prints

\bibitem[{{Ghosh}(1995)}]{ghosh95}
{Ghosh}, P. 1995, Journal of Astrophysics and Astronomy, 16, 289

\bibitem[{{Guti{\'e}rrez-Soto} {et~al.}(2011){Guti{\'e}rrez-Soto}, {Reig},
  {Fabregat}, \& {Fox-Machado}}]{juanguso11}
{Guti{\'e}rrez-Soto}, J., {Reig}, P., {Fabregat}, J., \& {Fox-Machado}, L.
  2011, in IAU Symposium, Vol. 272, IAU Symposium, ed. C.~{Neiner}, G.~{Wade},
  G.~{Meynet}, \& G.~{Peters}, 505--506

\bibitem[{{Hanuschik}(1989)}]{hanuschik89}
{Hanuschik}, R.~W. 1989, \apss, 161, 61

\bibitem[{{Heindl} {et~al.}(2001){Heindl}, {Coburn}, {Gruber}, {Rothschild},
  {Kreykenbohm}, {Wilms}, \& {Staubert}}]{heindl01}
{Heindl}, W.~A., {Coburn}, W., {Gruber}, D.~E., {et~al.} 2001, \apjl, 563, L35

\bibitem[{{Henrichs}(1982)}]{henrichs82}
{Henrichs}, H.~F. 1982, PhD thesis, University of Amsterdam

\bibitem[{{Henrichs}(1983)}]{henrichs83}
{Henrichs}, H.~F. 1983, in Accretion-Driven Stellar X-ray Sources, ed. W.~H.~G.
  {Lewin} \& E.~P.~J. {van den Heuvel}, 393--429

\bibitem[{{Illarionov} \& {Kompaneets}(1990)}]{illarionov&kompaneets90}
{Illarionov}, A.~F. \& {Kompaneets}, D.~A. 1990, \mnras, 247, 219

\bibitem[{{K{\i}z{\i}lo{\v g}lu} {et~al.}(2005){K{\i}z{\i}lo{\v g}lu},
  {K{\i}z{\i}lo{\v g}lu}, \& {Baykal}}]{kiziloglu05}
{K{\i}z{\i}lo{\v g}lu}, {\"U}., {K{\i}z{\i}lo{\v g}lu}, N., \& {Baykal}, A.
  2005, \aj, 130, 2766

\bibitem[{{K{\i}z{\i}lo{\v g}lu} {et~al.}(2007){K{\i}z{\i}lo{\v g}lu},
  {K{\i}z{\i}lo{\v g}lu}, {Baykal}, {Yerli}, \& {{\"O}zbey}}]{kiziloglu07}
{K{\i}z{\i}lo{\v g}lu}, U., {K{\i}z{\i}lo{\v g}lu}, N., {Baykal}, A., {Yerli},
  S.~K., \& {{\"O}zbey}, M. 2007, \aap, 470, 1023

\bibitem[{{Krimm} {et~al.}(2010){Krimm}, {Barthelmy}, {Baumgartner},
  {Cummings}, {Fenimore}, {Gehrels}, {Markwardt}, {Palmer}, {Sakamoto},
  {Skinner}, {Stamatikos}, {Tueller}, \& {Ukwatta}}]{krim10}
{Krimm}, H.~A., {Barthelmy}, S.~D., {Baumgartner}, W., {et~al.} 2010, The
  Astronomer's Telegram, 2663, 1

\bibitem[{{Krimm} {et~al.}(2013){Krimm}, {Holland}, {Corbet}, {Pearlman},
  {Romano}, {Kennea}, {Bloom}, {Barthelmy}, {Baumgartner}, {Cummings},
  {Gehrels}, {Lien}, {Markwardt}, {Palmer}, {Sakamoto}, {Stamatikos}, \&
  {Ukwatta}}]{krimm13}
{Krimm}, H.~A., {Holland}, S.~T., {Corbet}, R.~H.~D., {et~al.} 2013, \apjs,
  209, 14

\bibitem[{{Lattimer} (2012)}]{latti12}
{Lattimer}, J.~M., 2012, ARNPS, 62, 485

\bibitem[{{Lenz} \& {Breger}(2005)}]{Lenz05}
{Lenz}, P. \& {Breger}, M. 2005, Communications in Asteroseismology, 146, 53

\bibitem[{{Lyuty} \& {Zaitseva}(2000)}]{lyuty00}
{Lyuty}, V.~M. \& {Zaitseva}, G.~V. 2000, VizieR Online Data Catalog, 902,
  60013

\bibitem[{{M{\"u}ller} {et~al.}(2012){M{\"u}ller}, {K{\"u}hnel}, {Caballero},
  {Pottschmidt}, {F{\"u}rst}, {Kreykenbohm}, {Sagredo}, {Obst}, {Wilms},
  {Ferrigno}, {Rothschild}, \& {Staubert}}]{muller12}
{M{\"u}ller}, S., {K{\"u}hnel}, M., {Caballero}, I., {et~al.} 2012, \aap, 546,
  A125

\bibitem[{{M{\"u}ller} {et~al.}(2010){M{\"u}ller}, {K{\"u}hnel}, {Pottschmidt},
  {Caballero}, {F{\"u}rst}, {Barrag{\'a}n}, {Finger}, {Santangelo}, {Ferrigno},
  {Kreykenbohm}, \& {Wilms}}]{muller10}
{M{\"u}ller}, S., {K{\"u}hnel}, M., {Pottschmidt}, K., {et~al.} 2010, The
  Astronomer's Telegram, 3077, 1

\bibitem[{{Nakajima} {et~al.}(2010){Nakajima}, {Mihara}, {Nakagawa}, {Serino},
  {Sugizaki}, {Yamamoto}, {Sootome}, {Matsuoka}, {Negoro}, {Ozawa}, {Suwa},
  {Kawai}, {Morii}, {Sugimori}, {Usui}, {Ueno}, {Tomida}, {Kohama}, {Ishikawa},
  {Yoshida}, {Yamaoka}, {Nakahira}, {Tsunemi}, {Kimura}, {Kitayama}, {Ueda},
  {Isobe}, {Eguchi}, {Hiroi}, {Shidatsu}, {Matsumura}, {Yamazaki}, {Uzawa},
  {Daikyuji}, \& {Maxi Team}}]{nakajima10}
{Nakajima}, M., {Mihara}, T., {Nakagawa}, Y.~E., {et~al.} 2010, The
  Astronomer's Telegram, 3048, 1

\bibitem[{{Okazaki} {et~al.}(2013){Okazaki}, {Hayasaki}, \&
  {Moritani}}]{okazaki13}
{Okazaki}, A.~T., {Hayasaki}, K., \& {Moritani}, Y. 2013, \pasj, 65, 41

\bibitem[{{Papitto} {et~al.}(2012){Papitto}, {Torres}, \& {Rea}}]{papitto12}
{Papitto}, A., {Torres}, D.~F., \& {Rea}, N. 2012, \apj, 756, 188

\bibitem[{{Priedhorsky} \& {Holt}(1987)}]{priedholt87}
{Priedhorsky}, W.~C. \& {Holt}, S.~S. 1987, \ssr, 45, 291

\bibitem[{{Roberts} {et~al.}(1987){Roberts}, {Lehar}, \& {Dreher}}]{roberts87}
{Roberts}, D.~H., {Lehar}, J., \& {Dreher}, J.~W. 1987, \aj, 93, 968

\bibitem[{{Scargle}(1982)}]{scargle82}
{Scargle}, J.~D. 1982, \apj, 263, 835

\bibitem[{{Sharma}{et~al.}(2015){Sharma}, {Centelles}, {Vinas}, {Baldo}, {Burgio}}]{sharma15}
{Sharma}, B.~K., {Centelles}, M., {Vinas}, X., {Baldo}, {M.}, {Burgio}, G.~F. 2015, arXiv:1506.00375

\bibitem[{{Silaj} {et~al.}(2010){Silaj}, {Jones}, {Tycner}, {Sigut}, \&
  {Smith}}]{silaj10}
{Silaj}, J., {Jones}, C.~E., {Tycner}, C., {Sigut}, T.~A.~A., \& {Smith}, A.~D.
  2010, \apjs, 187, 228

\bibitem[{{Smith} \& {Takeshima}(1998)}]{smtake98}
{Smith}, D.~A. \& {Takeshima}, T. 1998, The Astronomer's Telegram, 36, 1

\bibitem[{{Stella} {et~al.}(1986){Stella}, {White}, \& {Rosner}}]{stella86}
{Stella}, L., {White}, N.~E., \& {Rosner}, R. 1986, \apj, 308, 669

\bibitem[{{Verrecchia} {et~al.}(2002{\natexlab{}}){Verrecchia}, {Israel},
  {Negueruela}, {Covino}, {Polcaro}, {Clark}, {Steele}, {Gualandi}, {Speziali},
  \& {Stella}}]{verrecchia02}
{Verrecchia}, F., {Israel}, G.~L., {Negueruela}, I., {et~al.}
  2002{\natexlab{}}, \aap, 393, 983

\bibitem[{{Wilson} {et~al.}(2003){Wilson}, {Finger}, {Coe}, \&
  {Negueruela}}]{wilson03}
{Wilson}, C.~A., {Finger}, M.~H., {Coe}, M.~J., \& {Negueruela}, I. 2003, \apj,
  584, 996

\bibitem[{{Wilson} {et~al.}(1998){Wilson}, {Finger}, {Wilson}, \&
  {Scott}}]{wilson981}
{Wilson}, C.~A., {Finger}, M.~H., {Wilson}, R.~B., \& {Scott}, D.~M. 1998,
  \iaucirc, 7014, 2

\end{thebibliography}

\end{document}